\documentclass[12pt,twoside]{article}
\usepackage{amsmath,a4wide,latexsym}
\usepackage{slashed}
\usepackage{bbm}
\usepackage{cite}
\normalsize
\begin{document}
\bibliographystyle{plain}
\begin{titlepage}
\begin{flushright}
UWThPh-2017-29 \\

\end{flushright}
\vspace{2.5cm}
\begin{center}
{\Large
Neutrino masses in a conformal multi-Higgs-doublet model
}
\\[40pt]
{\Large  Manuel Fink and Helmut Neufeld$^*$}

\vspace{1cm}

University of Vienna, Faculty of Physics,
Boltzmanngasse 5, A-1090 Wien, Austria

\vspace{1cm}


\vfill
{\bf Abstract} \\
\end{center}
\noindent
We construct a conformal version of a general multi-Higgs-doublet model with additional right-handed neutrino gauge-singlets.
Assuming a minimal extension of the scalar sector by a real singlet field, we show that the resulting model achieves the same attractive 
properties as the non-conformal theory, combining the seesaw mechanism and higher-order mass production to generate naturally light neutrino masses.
Starting with dimensionless couplings only, all masses and energy scales in the theory (including the heavy Majorana masses and the electroweak scale) 
are obtained 
from dimensional transmutation via the Coleman-Weinberg mechanism. 
A characteristic feature of the conformal model is the appearance 
of the ``scalon" in the scalar spectrum. The mass of this particle, which can be expressed in terms of 
the masses of the other particles in the theory, is produced at the one-loop level. We establish a connection between the large seesaw scale and a 
suppression of the scalon interactions. The positivity condition for the squared scalon mass requires sufficiently large masses of the additional Higgs 
bosons balancing the contributions of the heavy neutrinos.

\vfill
\noindent * Corresponding author (E-mail address: Helmut.Neufeld@univie.ac.at).

\end{titlepage}
\section{Introduction}
\label{sec: Introduction}
\renewcommand{\theequation}{\arabic{section}.\arabic{equation}}
\setcounter{equation}{0}

Already some time ago, W. Grimus and one of the present authors 
had suggested
a simple extension of the standard model (SM) with an arbitrary number $n_{\rm H}$ of
Higgs doublets $\Phi_k$ and $n_{\rm R}$ right-handed neutrino singlets $\nu_{\rm R}$ \cite{Grimus:1989pu}.
In this model, the latter become massive through an explicit Majorana mass term
\begin{equation}
\label{explicit-Majorana-mass-term}
- \frac{1}{2}
\left(
\overline{\nu_{\rm R}} M_{\rm R} (\nu_{\rm R})^{\rm c}
+
\overline{(\nu_{\rm R})^{\rm c}} M_{\rm R}^\ast \nu_{\rm R}
\right)
\end{equation}
and
interact with the usual $n_{\rm L} =3$ lepton doublets $L$ of the SM via the Yukawa couplings
\begin{equation}
\label{R-L-Yuk}
-
\sum\limits_{k=1}^{n_{\rm H}}
\left(
\widetilde{\Phi}_k^\dagger
\overline{{\nu}_{\rm R}} \, Y_{{\rm D} k}  L
+
\overline{L} Y_{{\rm D} k}^\dagger \nu_{\rm R}
\widetilde{\Phi}_k \right).
\end{equation}
In both (\ref{explicit-Majorana-mass-term}) and (\ref{R-L-Yuk}), a vector and matrix notation is employed, such that
$M_{\rm R}= M_{\rm R}^T$
is a complex
$
n_{\rm R}
\times
n_{\rm R}
$
matrix and the $n_{\rm H}$ complex matrices $Y_{{\rm D}k}$ are
$
n_{\rm R}
\times
n_{\rm L}
$.

The model contains two characteristic scales related to two distinct types of parameters with non-vanishing mass dimension. First,
the elements of the
Majorana mass matrix $M_{\rm R}$ (with their typical size denoted by $m_{\rm R}$) and second,
the entries $\mu^2_{ij}$ of the bilinear term in the scalar potential \cite{Grimus:2002ux}
\begin{equation}
\label{scalar-potential}
\sum\limits_{i,j} \mu^2_{ij} \Phi_i^\dagger \Phi_j
+
\sum\limits_{i,j,k,l} \lambda_{ijkl}(\Phi_i^\dagger
\Phi_j)(\Phi_k^\dagger \Phi_l).
\end{equation}
The parameters $\mu^2_{ij}$ and $\lambda_{ijkl}$
must be tuned in such a way that the potential assumes its minimum value at $\langle \Phi^0_k \rangle =
v_k/\sqrt{2}$ generating spontaneous symmetry breaking (SSB)
of the electroweak (EW) gauge group ${\rm SU}(2)_{\rm L} \times {\rm U}(1)_Y \to {\rm U}(1)_{\rm em}$ at the EW scale
\begin{equation}
v = \left( |v_1|^2 + \ldots +|v_{n_{\rm H}}|^2 \right)^{1/2} \simeq 246 \, {\rm GeV}.
\end{equation}
As a consequence, the Yukawa term (\ref{R-L-Yuk}) produces a Dirac mass term
\begin{equation}
\label{Dirac-mass-term}
- \overline{\nu_{\rm R}} M_{\rm D} \nu_{\rm L}
- \overline{\nu_{\rm L}} M_{\rm D}^\dagger \nu_{\rm R},
\qquad
M_{\rm D} =
\frac{1}{\sqrt{2}}
\sum\limits_{k=1}^{n_{\rm H}} v_k Y_{{\rm D} k},
\end{equation}
with its typical scale denoted by $m_{\rm D}$.

It was shown in  \cite{Grimus:1989pu} that for
$m_{\rm D} \ll m_{\rm R}$ and $n_{\rm R} < n_{\rm L}$, the model
combines the seesaw mechanism 
\cite{Minkowski:1977sc,Yanagida:1979as,Glashow:1979nm, GellMann:1980vs,Mohapatra:1979ia}
with higher-order mass production to generate naturally small Majorana neutrino
masses. Subsequently, a general analysis of the one-loop corrections to the seesaw
mechanism in this model was presented in \cite{Grimus:2002nk}.
Several phenomenological aspects of the simplest version of this model with just two Higgs doublets and one right-handed neutrino were discussed in
\cite{Grimus:1999wm} and more recently in \cite{Ibarra:2011gn}.

It is worth noting that only the terms with (operator) dimension two and three in the Lagrangian (i.e. the scalar mass term in
(\ref{scalar-potential}) and the Majorana mass term (\ref{explicit-Majorana-mass-term}), respectively)
spoil the conformal invariance of the action of
the model.

The situation is quite similar in the SM, where only the term bilinear in the Higgs doublet field (responsible for the EW scale) violates
conformal symmetry explicitly. The introduction of an explicit scalar mass-term in the SM
does not only break conformal invariance, but it also gives rise to the
so-called hierarchy problem. In this context, Bardeen
\cite{Bardeen:1995kv} emphasized
the importance of the protective property of classical scale
invariance (softly broken by the Higgs mass term) for the fine-tuning problem of the SM.
Taking this perspective, the popular bugbear of frightening quadratic divergences boils down to the mere artefact of an inadequate regularization,
taming the quadratic sensitivity of quantum corrections to high mass scales to a logarithmic one.

Going one step further, it is tempting to assume that (as in QCD) nature abhors the presence of dimensionful couplings in the Lagrangian of a
fundamental theory and the generation of mass scales is a pure quantum effect. This idea was put forward in the seminal work of
Coleman and E. Weinberg (CW)
\cite{Coleman:1973jx}. By employing the one-loop effective potential \cite{Goldstone:1962es,JonaLasinio:1964cw} as a convenient tool, CW demonstrated 
the breaking  of
classical scale invariance by higher order quantum
corrections and the generation of a mass scale by dimensional transmutation \cite{Coleman:1973jx}.
Subsequently, the most efficient method to investigate the one-loop effective potential of general conformal gauge theories was developed by Gildener
and S. Weinberg (GW) \cite{Gildener:1976ih},
which allows one to easily disentangle leading and subleading contributions.

A common property of such models is the appearance of the ``scalon" $S$ \cite{Gildener:1976ih}, a scalar state with universal couplings and a one-loop
mass $M_S$, which can be expressed in terms of the masses of the other particles present in the theory by the relation
\begin{equation}
\label{scalon-mass-general}
M_{S}^2 = \frac{1}{8 \pi^2 V^2} \left( \sum\limits_s M_s^4 + 3 \sum\limits_g M_g^4 - 2 \sum\limits_f m_f^4 \right).
\end{equation}
The indices $s$, $g$ and $f$ enumerate the scalars, gauge bosons and
Weyl fermions (two-component spinors), respectively. The quantity $V$ is a vacuum expectation value  characterizing the scale of
the model generated by dimensional transmutation \cite{Coleman:1973jx}.

Note that the positivity of (\ref{scalon-mass-general}) is the criterion for the existence of a
minimum of the effective potential. This shows in hindsight why early attempts \cite{Gildener:1976ih}
to interpret the SM Higgs  as the scalon of a conformal
version of the SM (without enlarging its particle content) were doomed to fail: the relevant term $3 M_Z^4 + 6 M_W^4 - 12 m_t^4$ is negative because of
the large mass of the top quark.

Motivated by several current puzzles in particle physics,
the last few years have seen a strong revival of interest in conformal extensions of the SM (see for instance \cite{Helmboldt:2016mpi}
and the references therein). It is a common feature of such models that, due to (\ref{scalon-mass-general}), heavy fermionic states (like the top quark
or heavy Majorana neutrinos) must be balanced by the presence of bosonic states to meet the positivity requirement $M_S^2 > 0$. At the same time, all
mass scales (including the EW scale) are ``explained" by dimensional transmutation instead of through explicit mass terms.

In this work, we propose a conformal version of the multi-Higgs-doublet model with right-handed neutrinos.
In such a scenario, the explicit mass term
(\ref{explicit-Majorana-mass-term}) is forbidden and masses for the
right-handed neutrinos
can only be generated from Yukawa couplings \cite{Lindner:2014oea} after
SSB.
Restricting ourselves to a minimal extension of the particle content postulated in
\cite{Grimus:1989pu},
we introduce an additional real scalar singlet field $\varphi_0$, which interacts with
$\nu_{\rm R}$ via the Yukawa couplings
\begin{equation}
\label{Yukawa-coupling}
- \frac{\varphi_0}{2}
\left(
\overline{\nu_{\rm R}} \, Y_{\rm R} (\nu_{\rm R})^{\rm c}
+
\overline{(\nu_{\rm R})^{\rm c}} \, Y_{\rm R}^\ast \nu_{\rm R}
\right),
\end{equation}
where $Y_{\rm R} = Y_{\rm R}^T$ is a symmetric complex $n_{\rm R} \times n_{\rm R}$ matrix.
The Majorana mass matrix $M_{\rm R} = \langle \varphi_0 \rangle Y_{\rm R}$ is now produced through the vacuum expectation value of the singlet
scalar.

The characteristic scale $V$ of the model is the common origin of the vacuum expectation values of the neutral
scalars via
\begin{equation}
\label{vevs}
\langle \varphi_0 \rangle = \underbrace{n_0 V}_{v_0}, \qquad
\langle \Phi_k^0 \rangle = \underbrace{n_k V}_{v_k} / \sqrt{2}
\end{equation}
with coefficients
\begin{equation}
n_0 > 0, \quad n_k \in \mathbbm{C}, \qquad
n_0^2 + \sum\limits_{k=1}^{n_{\rm H}} n_k^\ast n_k =1,
\end{equation}
generating eventually all masses in the theory. This is in contrast to the non-conformal version of the model, where the origin of the EW
scale $v$ and the (heavy) Majorana scale $m_{\rm R}$ are completely unrelated.
Except for unnaturally small Yukawa couplings $Y_{{\rm D} k}$, the seesaw mass-hierarchy $m_{\rm D} \ll m_{\rm R}$ requires $|v_k| \ll v_0$ or, 
equivalently, $v \ll V$, which amounts to assuming
\begin{equation}
\label{inequ-n}
\sum\limits_{k=1}^{n_{\rm H}} n_k^\ast n_k 
\ll
n_0^2 
\simeq 1.
\end{equation}

In this work, we study the generic case of the proposed model with an arbitrary number $n_{\rm H}$ of Higgs doublets and a likewise unspecified 
number $n_{\rm R}$ of right-handed neutrino fields. Apart from gauge and conformal symmetries, no further symmetry constraints will be imposed.

The paper is organized along the following lines. 
In section \ref{sec: Description of the model} we describe the particle content of the model and construct the
conformal Lagrangian. 
The relevant aspects of SSB via the CW mechanism are addressed in section \ref{sec: SSB}. 
The tree-level mass matrices and the associated mass
eigenfields are
determined in section \ref{sec: Tree-level masses}. 
In section \ref{sec: Interaction terms} we list the electroweak gauge interactions of the fermions and the Yukawa couplings  
expressed in terms of mass eigenfields. 
The properties of the scalon are discussed in section \ref{sec: scalon}, where 
we establish a connection between the large seesaw scale and the suppression of the strength of the scalon couplings
and we also identify the experimentally observed $125 \, \rm GeV$ Higgs particle in the context of our model. 
The calculation of the one-loop neutrino masses is carried out in section \ref{sec: Neutrino masses at one-loop}  
elucidating the differences (and similarities) to the original non-conformal version of the multi-Higgs-doublet model.
In section \ref{sec: Conclusions} we present the final results and an outlook.

\section{Description of the model}
\label{sec: Description of the model}
\renewcommand{\theequation}{\arabic{section}.\arabic{equation}}
\setcounter{equation}{0}

We consider a classically scale-invariant
${\rm SU}(3)_{\rm c} \times {\rm SU}(2)_{\rm L} \times {\rm U}(1)_Y$
gauge theory
with a real scalar singlet field
and an arbitrary number $n_{\rm H}$ of scalar doublets,
\begin{equation}
\varphi_0, \qquad
\Phi_k =  \begin{pmatrix} \Phi_k^+ \\
                          \Phi_k^0 \\
                         \end{pmatrix}, \quad 1 \le k \le n_{\rm H}.
\end{equation}
The lepton sector consists of the usual SM fields
\begin{equation}
L =  \begin{pmatrix} \nu_{\rm L} \\
                     \ell_{\rm L} \\
                         \end{pmatrix}_{\! \! r}, \quad \ell_{{\rm R} r},  \quad 1 \le r \le n_{\rm L}=3,
\end{equation}
furnished in addition with $n_{\rm R}$ right-handed neutrino singlets,
\begin{equation}
                         \nu_{{\rm R} s},  \quad 1 \le s \le n_{\rm R}.
\end{equation}
The quark fields (transforming as triplets with respect to $\rm SU(3)_c$) are denoted by
\begin{equation}
Q =  \begin{pmatrix} u_{\rm L} \\
                     d_{\rm L}^\prime \\
                         \end{pmatrix}_{\! \! r}, \quad
u_{{\rm R}  r},  \quad
d_{{\rm R}  r},  \quad
1 \le r \le n_{\rm L}.
\end{equation}
The transformation properties of these fields
with respect to weak isospin and weak hypercharge $(T, Y/2)$ are given by
$\varphi_0 \sim (0,0)$,
$\Phi_k \sim (1/2, 1/2)$,
$L_r \sim (1/2, -1/2)$, $\ell_{{\rm R}  r} \sim
(0,-1)$, $\nu_{{\rm R}  s} \sim (0,0)$, $Q_r \sim (1/2, 1/6)$, $u_{{\rm R}  r} \sim (0,2/3)$, $d_{{\rm R}  r} \sim (0,-1/3)$.

The covariant derivative has the generic form
\begin{equation}
\label{cov der}
D_\mu = \partial_\mu + i g_{\rm s} \sum\limits_{a=1}^8 \mathcal{T}_a G^a_\mu + ig \vec{T} \cdot \vec{W}_\mu + i g^\prime \frac{Y}{2} B_\mu
\end{equation}
with
\begin{equation}
\mathcal{T}_a = \begin{cases}
                \lambda_a/2 & \text{for SU(3) triplets} \\
                0 & \text{for SU(3) singlets} \\
                \end{cases}, \qquad
\vec{T} = \begin{cases}
                \vec{\tau} /2 & \text{for SU(2) doublets} \\
                0 & \text{for SU(2) singlets} \\
                \end{cases},
\end{equation}
where the $\lambda_a$ 
and $\vec{\tau}$ 
denote the Gell-Mann and 
Pauli matrices, respectively.

With these building blocks, the construction of the Lagrangian is now straightforward. The gauge boson part reads
\begin{equation}
\mathcal{L}_{\rm g} = -
\frac{1}{4}
\sum\limits_{a=1}^8
G^a_{\mu \nu} G_a^{\mu \nu}
              -\frac{1}{4} \vec{W}_{\mu \nu} \vec{W}^{\mu \nu}
              -\frac{1}{4} B_{\mu \nu} B^{\mu \nu}
\end{equation}
and the fermionic Lagrangian is given by
\begin{equation}
\label{Lfermion}
\mathcal{L}_{\rm f} = \overline{L} i \slashed{D} L
+ \overline{{\ell}_{\rm R}} i \slashed{D} \ell_{\rm R}
+ \overline{{\nu}_{\rm R}} i \slashed{\partial} \nu_{\rm R}
+ \overline{Q} i \slashed{D} Q
+ \overline{{u}_{\rm R}} i \slashed{D} u_{\rm R}
+ \overline{{d}_{\rm R}} i \slashed{D} d_{\rm R},
\end{equation}
where the various field multiplets are now written as vectors in flavour space.
The kinetic and gauge interaction terms of the scalars are contained in
\begin{equation}
\label{L-s}
\mathcal{L}_{\rm s} = \frac{1}{2} \partial_\mu \varphi_0 \partial^\mu \varphi_0
+ \sum\limits_{k=1}^{n_{\rm H}} (D_\mu \Phi_k)^\dagger D^\mu \Phi_k.
\end{equation}

The Yukawa Lagrangian has the general form
\begin{equation}
\mathcal{L}_{\rm Y} =
- \frac{\varphi_0}{2}
\overline{\nu_{\rm R}}
\, Y_{\rm R}
(\nu_{\rm R})^{\rm c}
- \sum\limits_{k=1}^{n_{\rm H}} \left[ \left( \Phi_k^\dagger \overline{{\ell}_{\rm R}} Y^{(\ell)}_k+
\widetilde{\Phi}_k^\dagger \overline{{\nu}_{\rm R}} Y_{{\rm D} k} \right) L
+ \left( \Phi_k^\dagger \overline{{d}_{\rm R}} Y^{(d)}_ k+
\widetilde{\Phi}_k^\dagger \overline{{u}_{\rm R}} Y^{(u)}_k \right) Q \right]
+ {\rm h.c.}
\end{equation}
with
\begin{equation}
\widetilde{\Phi}_k = i \tau_2 \Phi_k^\ast =
\begin{pmatrix}
\Phi^{0 \ast}_k \\
- \Phi^{+ \ast}_k \\
\end{pmatrix}
\equiv
\begin{pmatrix}
\Phi^{0 \ast}_k \\
- \Phi^{-}_k \\
\end{pmatrix}
\end{equation}
and complex
$n_{\rm L} \times n_{\rm L}$ matrices
$Y^{(\ell)}_k, Y^{(u)}_k, Y^{(d)}_k$. 
The matrices
$Y_{{\rm D} k}$, $Y_{\rm R}$ had already been introduced in
(\ref{R-L-Yuk})
and
(\ref{Yukawa-coupling}), respectively.

Finally, the scalar potential is given by
\begin{equation}
\label{treepotential}
\mathcal{V}_0(\varphi_0, \Phi_k) =
\frac{\lambda_0}{4} \varphi_0^4
+
\sum\limits_{i,j,k,l} \lambda_{ijkl}(\Phi_i^\dagger
\Phi_j)(\Phi_k^\dagger \Phi_l) +
\varphi_0^2 \sum\limits_{i,j} \kappa_{ij} \Phi_i^\dagger \Phi_j,
\end{equation}
where hermiticity of $\mathcal{V}_0$ and the symmetry of the second term imply the relations
\begin{equation} 
\lambda_0 = \lambda_0^\ast, \, \,
\lambda_{ijkl} =\lambda_{klij}, \, \,
\lambda_{ijkl} =\lambda_{jilk}^\ast, \, \,
\kappa_{ij} = \kappa_{ji}^\ast.
\end{equation}

The total Lagrangian
\begin{equation}
\mathcal{L} =
\mathcal{L}_{\rm g}
+ \mathcal{L}_{\rm f}
+ \mathcal{L}_{\rm s}
+ \mathcal{L}_{\rm Y}
-\mathcal{V}_0
\end{equation}
contains all conformally invariant and gauge-symmetric terms that
can be composed of the given particle multiplets. Note that
the (explicit) mass terms
\begin{equation}
- \frac{1}{2} \overline{\nu_{\rm R}} M_{\rm R} (\nu_{\rm R})^{\rm c}
- \frac{1}{2} \overline{(\nu_{\rm R})^{\rm c}} M_{\rm R}^\ast \nu_{\rm R}
- \frac{\mu_0^2}{2} \varphi_0^2
-\sum\limits_{i,j} \mu^2_{ij} \Phi_i^\dagger \Phi_j
\end{equation}
are forbidden by scale invariance.

\section{Spontaneous symmetry breaking}
\label{sec: SSB}
\renewcommand{\theequation}{\arabic{section}.\arabic{equation}}
\setcounter{equation}{0}

Although the model appears to have a symmetric vacuum at tree level, scale invariance and gauge
symmetry are actually broken as a result of higher order quantum corrections via the
CW mechanism \cite{Coleman:1973jx}. A convenient tool to study the vacuum
structure of the theory is the effective potential \cite{Goldstone:1962es,JonaLasinio:1964cw}
\begin{equation}
\label{eff-pot}
\mathcal{V}(\varphi_0, \Phi_k) =
\mathcal{V}_0(\varphi_0, \Phi_k) + \delta \mathcal{V}(\varphi_0, \Phi_k),
\end{equation}
where $\delta \mathcal{V}$
contains the quantum corrections
to the tree-level potential $\mathcal{V}_0$.\footnote{The general form of $\delta \mathcal{V}$ in the Landau gauge at one-loop order was obtained
by CW \cite{Coleman:1973jx} using a specific renormalization scheme. The result was subsequently confirmed in \cite{Weinberg:1973ua}
using a different method. See e.g. \cite{Espinosa:2007qk}
for the corresponding formula in the $\rm \overline{MS}$ scheme.}

Following the GW approach \cite{Gildener:1976ih}, we consider the constant field configuration
\begin{equation}
\varphi_0 = N_0 \in \mathbbm{R}, \quad \Phi_k = \frac{1}{\sqrt{2}} N_k \in \mathbbm{C}^2
\end{equation}
subject to the constraint
\begin{equation}
N_0^2 + \sum\limits_{k=1}^{n_{\rm H}} N_k^\dagger N_k =1,
\end{equation}
which parametrizes the $4 n_{\rm H}$ dimensional unit sphere
${\rm S}^{4 n_{\rm H}}$.
We are looking for a minimum of the tree potential (\ref{treepotential})
on ${\rm S}^{4 n_{\rm H}}$ at a point
\begin{equation}
N_0 = n_0 > 0, \quad N_k = \begin{pmatrix} 0 \\
                                           n_k \\
                           \end{pmatrix}, \, \,  n_k \in \mathbbm{C}, \quad n_0^2 + \sum\limits_{k=1}^{n_{\rm H}} n_k^\ast n_k =1,
\end{equation}
so that the electromagnetic charge $Q_{\rm em} = T_3 + Y/2$ remains unbroken. In this way, we obtain the stationarity condition
\begin{equation}
\label{stat-cond}
\sum\limits_{j,k,l} \lambda_{ijkl} n_j n_k^\ast n_l + n_0^2 \sum\limits_j\kappa_{ij} n_j - n_i \left( \sum\limits_{k,l} \kappa_{kl} n_k^\ast n_l +
\lambda_0
n_0^2 \right) = 0 .
\end{equation}
In the next step we take advantage of the fact that the parameters of the tree potential are functions of the renormalization scale $\mu$,
\begin{equation}
\lambda_0 = \lambda_0(\mu), \, \, \lambda_{ijkl} = \lambda_{ijkl} (\mu), \, \, \kappa_{ij} = \kappa_{ij}(\mu),
\end{equation}
governed by the pertinent renormalization group equations. As suggested by GW
\cite{Gildener:1976ih},
we adjust the renormalization scale $\mu =
\Lambda_{\rm GW}$ in such a way that the minimum value of $\mathcal{V}_0$ on the unit sphere vanishes:
\begin{equation}
\label{GW-condition}
\mathcal{V}_0 \left( n_0,  \frac{n_k}{\sqrt{2}} \begin{pmatrix} 0 \\
                                                      1 \\
                                      \end{pmatrix} \right) =0.
\end{equation}
In our model, the GW condition (\ref{GW-condition})
corresponds to the equation
\begin{equation}
\label{GW-mod}
\sum\limits_{i,j,k,l} \lambda_{ijkl} n_i^\ast n_j n_k^\ast n_l + 2 n_0^2 \sum\limits_{i,j} \kappa_{ij} n_i^\ast n_j + \lambda_0 n_0^4 =0,
\end{equation}
which allows us to trade one of the couplings, say $\lambda_0$, for the GW scale $\Lambda_{\rm GW}$, being an example of dimensional
transmutation \cite{Coleman:1973jx}. In the following, we shall assume that all couplings of our model are taken at $\mu = \Lambda_{\rm GW}$ so that
(\ref{GW-mod}) holds. Note also that (\ref{stat-cond}) and (\ref{GW-mod}) are equivalent to the system of equations
\begin{eqnarray}
 \sum\limits_j \left( \sum\limits_{k,l} \lambda_{ijkl} n_k^\ast n_l + n_0^2 \kappa_{ij} \right) n_j &=& 0, \label{eq1} \\
 \sum\limits_{i,j} \kappa_{ij} n_i^\ast n_j + \lambda_0 n_0^2 &=& 0. \label{eq2}
\end{eqnarray}

As the tree potential is a homogeneous function of the field variables,
$\mathcal{V}_0(\varphi_0, \Phi_k)$ attains a minimum value zero everywhere on the ray
\begin{equation}
\label{ray}
\varphi_0 = \phi n_0, \quad \Phi_k = \frac{\phi n_k}{\sqrt{2}} \begin{pmatrix} 0 \\
                                                                            1 \\
                                                            \end{pmatrix}, \quad \phi \in \mathbbm{R},
\end{equation}
once (\ref{GW-condition}) holds. When the higher-order term $\delta \mathcal{V}(\varphi_0, \Phi_k)$ in (\ref{eff-pot}) is turned on, the effective
potential receives a
small curvature in the radial direction, which picks out a definite value $\langle \phi \rangle = V > 0$ and also a small shift at this minimum
\cite{Gildener:1976ih}:
\begin{equation}
\label{vacuum}
\left\langle \varphi_0 \right\rangle = \underbrace{V n_0}_{v_0} + \delta v_0, \quad \left\langle \Phi_k^0 \right\rangle =
\frac{1}{\sqrt{2}}
(\underbrace{V n_k}_{v_k} +
\delta v_k).
\end{equation}

\section{Tree-level masses}
\label{sec: Tree-level masses}
\renewcommand{\theequation}{\arabic{section}.\arabic{equation}}
\setcounter{equation}{0}

The lowest-order mass matrices of the model are obtained by expanding the scalar fields around the leading terms in (\ref{vacuum}),
\begin{equation}
\label{def-rho-sigma}
\varphi_0 = v_0 + \rho_0, \quad
\Phi_k = \begin{pmatrix} \Phi_k^+ \\
\frac{1}{\sqrt{2}} (v_k + \rho_k + i \sigma_k) \\
\end{pmatrix}
\end{equation}
with real fields $\rho_0$, $\rho_k$ and $\sigma_k$.

\subsection{Gauge bosons}
\label{subsec: Gauge bosons}

From $\mathcal{L}_{\rm s}$ one obtains the mass eigenvalues of the vector bosons,
\begin{equation}
M_W^2= v^2 g^2/4,
\quad M_Z^2 = v^2 (g^2 + g^{\prime \, 2})/4,
\quad v^2 = \sum\limits_{k=1}^{n_{\rm H}} v_k^\ast v_k,
\end{equation}
with $v = (\sqrt{2} \, G_{\rm F})^{-1/2} \simeq 246 \, {\rm GeV}$.
The associated
mass eigenfields are given by
\begin{equation}
W_\mu^\pm = \frac{(W_1 \mp i W_2)_\mu}{\sqrt{2}}, \quad
\begin{pmatrix} Z_\mu \\
                A_\mu \\
\end{pmatrix}
=
\begin{pmatrix}
\cos \theta_{\rm W} & - \sin \theta_{\rm W} \\
\sin \theta_{\rm W} &   \cos \theta_{\rm W} \\
\end{pmatrix}
\begin{pmatrix} W_{3 \mu} \\
                     B_\mu \\
\end{pmatrix},
\end{equation}
with the Weinberg angle $\theta_{\rm W}$ defined by
$
\sin \theta_{\rm W} = g^\prime / \sqrt{g^2 + g^{\prime \, 2}}
$. The $Z^0$ mass formula can now be rewritten as 
\begin{equation}
M_Z^2 = \frac{g^2 v^2}{4 \cos^2 \theta_{\rm W}}.
\end{equation}

\subsection{Charged scalars}
\label{susec: Charged scalars}

The mass matrix of the charged scalars (derived from $\mathcal{V}_0$) reads
\begin{equation}
(M_+^2)_{ij} = V^2 \left( \sum\limits_{k, l} \lambda_{i j k l}n_k^\ast n_l + n_0^2 \kappa_{ij} \right).
\end{equation}
The relation (\ref{eq1}) is equivalent to the eigenvalue equation
\begin{equation}
M_+^2 n =0, \quad
n = \begin{pmatrix} n_1  \ldots  n_{n_{\rm H}}\\ \end{pmatrix}^T,
\end{equation}
which means that the vector
$n$ is an eigenvector of the hermitian matrix $M_+^2$ associated with the eigenvalue zero.

The fields $\Phi_k^+$ are related to the mass eigenfields $H_a^+$ by a unitary transformation $U$,
\begin{equation}
\Phi_k^+ = \sum\limits_{a=1}^{n_{\rm H}} U_{k a} H_a^+, \quad
U^\dagger M_+^2 U = {\rm diag} \begin{pmatrix} \mu_1^2, \ldots,  \mu_{n_{\rm H}-1}^2, 0 \\ \end{pmatrix}, \quad
U^\dagger = U^{-1}.
\end{equation}
The model contains $n_{\rm H} -1$ physical charged
scalars $H_1^\pm, \ldots, \,  H^\pm_{n_{\rm H}-1}$ and one
charged
would-be-Goldstone-boson $H^\pm_{n_{\rm H}} \equiv G^\pm$ corresponding to the last column of the unitary matrix $U$ with
\begin{equation}
\label{U-k-nH}
U_{k \, n_{\rm H}}= \frac{n_k}{\sqrt{1-n_0^2}}.
\end{equation}

\subsection{Neutral scalars}
\label{subsec: Neutral scalars}

The tree-level mass term of the neutral scalars can be written in the form
\begin{equation}
- \frac{1}{2} \begin{pmatrix} \rho_0 & \rho^T & \sigma^T \\ \end{pmatrix}
\underbrace{\begin{pmatrix} 2 \lambda_0 v_0^2 & {\rm Re} \, k^T & {\rm Im} \, k^T \\
                {\rm Re} \, k        &      A       &      C       \\
                {\rm Im} \, k        &      C^T     &      B       \\
\end{pmatrix}}_{\mathcal{M}_0^2}
\begin{pmatrix} \rho_0 \\
                  \rho \\
                  \sigma \\
\end{pmatrix},
\end{equation}
with
$
\rho^T = \begin{pmatrix} \rho_1 \ldots  \rho_{n_{\rm H}}\\
         \end{pmatrix}$ 
and $\sigma^T = \begin{pmatrix} \sigma_1  \ldots
                      \sigma_{n_{\rm H}}\\
         \end{pmatrix}$.
The vector $k \in \mathbbm{C}^{n_{\rm H}}$ is defined by
\begin{equation}
k_i = 2 V^2 n_0  \sum\limits_{j=1}^{n_{\rm H}} \kappa_{ij} n_j.
\end{equation}
The $n_{\rm H} \times n_{\rm H}$ matrices
\begin{equation}
A = {\rm Re} \, (M_+^2 +K^\prime +K), \quad
B = {\rm Re} \, (M_+^2 +K^\prime -K), \quad
C = {\rm Im} \, (- M_+^2  - K^\prime +K)
\end{equation}
can be written in terms of $M_+^2$ and the matrices $K=K^T$, $K^\prime = K^{\prime \, \dagger}$ introduced in \cite{Grimus:2002ux}:
\begin{equation}
K_{ij} = V^2 \sum\limits_{k,l} \lambda_{ikjl} n_k n_l, \quad
K_{ij}^\prime = V^2 \sum\limits_{k,l} \lambda_{iklj} n_k n_l^\ast.
\end{equation}

The two orthogonal vectors
\begin{equation}
\label{eigenvectors}
\begin{pmatrix} n_0 \\
                {\rm Re} \, n \\
                {\rm Im} \, n \\
\end{pmatrix},
\, \,
\begin{pmatrix}  0\\
                - {\rm Im} \, n \\
                {\rm Re} \, n \\
\end{pmatrix}
\in \mathbbm{R}^{2 n_{\rm H} +1}
\end{equation}
are eigenvectors of $\mathcal{M}_0^2$ with eigenvalue zero.  The first one is related to the scalon $S$ \cite{Gildener:1976ih}, which receives a
nonvanishing
mass only at one-loop order. The second one is associated with the neutral would-be-Goldstone-boson $G^0$.

The neutral scalar mass eigenfields
$S^0_b$ ($b=0, \ldots, \,  2n_{\rm H}$)
are related
to $\rho_0$, $\rho$ and $\sigma$ by
an orthogonal
$(2 n_{\rm H}+1) \times (2 n_{\rm H}+1)$ matrix
$\mathcal{R}$,
\begin{equation}
\begin{pmatrix} \rho_0 \\
                  \rho \\
                  \sigma \\
\end{pmatrix}
=
\underbrace{\begin{pmatrix} r^T \\
                {\rm Re} \, R \\
                {\rm Im} \, R \\
\end{pmatrix}}_{\mathcal{R}}
\begin{pmatrix} S^0_0 \\
                \cdots \\
                 S^0_{2 n_{\rm H}} \\
\end{pmatrix}
\end{equation}
with a complex $n_{\rm H} \times (2 n_{\rm H} +1)$ matrix
$R = {\rm Re} \, R + i \, {\rm Im} \, R$ and
$r \in \mathbbm{R}^{2 n_{\rm H} +1}$. In index notation, we write
$
\mathcal{R}_{0  b} = r_b$,
$\mathcal{R}_{k  b} = {\rm Re} \, R_{k  b}$,
$\mathcal{R}_{(k+n_{\rm H})  b} = {\rm Im} \, R_{k  b}$ with
$1 \le k \le n_{\rm H}$ and $0 \le b \le 2 n_{\rm H}$.

The orthogonality of $\mathcal{R}$ can be expressed in the form
\begin{equation}
r r^T  + {\rm Re} \, (R^\dagger R) = \mathbbm{1}_{2 n_{\rm H} +1},
\end{equation}
or, equivalently, by the relations
\begin{equation}
\label{orth-rel}
r^T r=1, \quad Rr=0, \quad
R R^\dagger = 2 \cdot \mathbbm{1}_{n_{\rm H}}, \quad
R R^T = 0.
\end{equation}

The numbering of the mass eigenfields is chosen in such a way that $S^0_0 \equiv S$ and $S^0_{2 n_{\rm H}} \equiv G^0$ so that
\begin{equation}
\mathcal{R}^T \mathcal{M}_0^2 \, \mathcal{R} = {\rm diag} \begin{pmatrix} 0, \, M_1^2, \ldots, \, M_{2 n_{\rm H}-1}^2, \, 0 \end{pmatrix}.
\end{equation}
As a consequence of
(\ref{eigenvectors}), the
first and the last column of $\mathcal{R}$ are then obtained from
$r_0 = n_0$, $R_{k  0} = n_k$ (for $b=0$) and
$r_{2 n_{\rm H}} = 0$, $R_{k \, 2n_{\rm H}} = i n_k /{\sqrt{1-n_0^2}}$ (for
$b= 2 n_{\rm H}$),
which implies
\begin{equation}
\label{rel-rho-sigma}
\rho_0 = n_0 \, S + \sum\limits_{b=1}^{2 n_{\rm H}-1} \! r_b \, S^0_b, \quad
\rho_k + i \sigma_k = n_k \, S + \sum\limits_{b=1}^{2 n_{\rm H}-1} \! R_{k  b}  \, S^0_b
+ \frac{i n_k}{\sqrt{1-n_0^2}} \, G^0.
\end{equation}

\subsection{Fermions}
\label{subsec: Fermions}

Without loss of generality, we may assume that $\ell_{\rm L,R}$, $u_{\rm L,R}$ and $d_{\rm R}$ are already mass eigenfields, which can always be
achieved by a suitable basis transformation in flavour space. The weak eigenfield $d^\prime_{\rm L} = V_{\rm CKM} d_{\rm L}$ is related to the mass
eigenfield $d_{\rm L}$ by the weak mixing matrix $V_{\rm CKM}$ \cite{Cabibbo:1963yz,Kobayashi:1973fv}. Defining the Dirac fields
$\ell = \ell_{\rm L} + \ell_{\rm R}$, $u = u_{\rm L} + u_{\rm R}$, $d = d_{\rm L} + d_{\rm R}$, the fermion mass terms are given by
\begin{equation}
- \overline{\ell} M_\ell \ell
- \overline{u} M_u u
- \overline{d} M_d d
- \overline{\nu_{\rm R}} M_{\rm D} \nu_{\rm L}
- \overline{\nu_{\rm L}} M_{\rm D}^\dagger \nu_{\rm R}
- \frac{1}{2} \overline{\nu_{\rm R}} M_{\rm R} (\nu_{\rm R})^{\rm c}
- \frac{1}{2} \overline{(\nu_{\rm R})^{\rm c}} M_{\rm R}^\ast \nu_{\rm R},
\end{equation}
where
\begin{eqnarray}
M_\ell &=& \frac{1}{\sqrt{2}} \sum\limits_{k=1}^{n_{\rm H}} v_k^\ast Y^{(\ell)}_k = {\rm diag} \begin{pmatrix} m_e, \, m_\mu, \, m_\tau 
\end{pmatrix}, 
\label{M-l} \\
M_u &=& \frac{1}{\sqrt{2}} \sum\limits_{k=1}^{n_{\rm H}} v_k Y^{(u)}_k = {\rm diag} \begin{pmatrix} m_u, \, m_c, \, m_t \end{pmatrix}, 
\label{M-u} \\
M_d &=& \frac{1}{\sqrt{2}} \sum\limits_{k=1}^{n_{\rm H}} v_k^\ast Y^{(d)}_k V_{\rm CKM} = {\rm diag} \begin{pmatrix} m_d, \, m_s, \, m_b 
\end{pmatrix},
\label{M-d} \\
M_{\rm D} &=& \frac{1}{\sqrt{2}} \sum\limits_{k=1}^{n_{\rm H}} v_k Y_{{\rm D} k},
\quad M_{\rm R} = v_0 Y_{\rm R}.
\label{M-D}
\end{eqnarray}

Introducing the left-handed field
\begin{equation}
\omega_{\rm L} = \begin{pmatrix} \nu_{\rm L} \\
                                 (\nu_{\rm R})^{\rm c} \\
                 \end{pmatrix},
\end{equation}
the neutrino mass term
can be written in the compact form
\begin{equation}
- \frac{1}{2} \overline{(\omega_{\rm L})^{\rm c}} \mathcal{M}_\nu \, \omega_{\rm L} + {\rm h.c.},
\end{equation}
with an $(n_{\rm L} + n_{\rm R})  \times (n_{\rm L} + n_{\rm R})$ symmetric mass matrix
\begin{equation}
\label{M-nu}
\mathcal{M}_\nu
= \begin{pmatrix} 0 & M_{\rm D}^T \\
                                  M_{\rm D} & M_{\rm R} \\
                  \end{pmatrix}.
\end{equation}
To obtain the (tree-level) mass eigenfields
$\hat{\omega}_{\rm L}$,
we perform a unitary transformation
\begin{equation}
\label{intro-U}
\omega_{\rm L} =
\mathcal{U}
\hat{\omega}_{\rm L}, \quad
\mathcal{U}^\dagger
=
\mathcal{U}^{-1},
\end{equation}
such that
\begin{equation}
\label{diag-prop}
\mathcal{U}^T
\mathcal{M}_\nu \,
\mathcal{U}
=
\hat{\mathcal{M}}_\nu
\end{equation}
is diagonal and non-negative.

It is convenient to decompose the unitary
$
(n_{\rm L} + n_{\rm R}) \times
(n_{\rm L} + n_{\rm R})
$ matrix
$
\mathcal{U}
$
as \cite{Grimus:1989pu}
\begin{equation}
\label{decomp-1}
\mathcal{U}
= \begin{pmatrix} U_{\rm L} \\
                  U_{\rm R}^\ast \\
  \end{pmatrix}
\end{equation}
with an
$
n_{\rm L} \times
(n_{\rm L} + n_{\rm R})
$
submatrix $U_{\rm L}$ and an
$
n_{\rm R} \times
(n_{\rm L} + n_{\rm R})
$
submatrix $U_{\rm R}$. Defining
\begin{equation}
\chi =
\hat{\omega}_{\rm L}
+
(\hat{\omega}_{\rm L})^{\rm c},
\end{equation}
the weak eigenfields
$\nu_{\rm L,R}$
can be written as linear combinations of the neutrino mass-eigenfields
\begin{equation}
\nu_{\rm L} =
U_{\rm L} \hat{\omega}_{\rm L} =
U_{\rm L} P_{\rm L} \chi,
\quad
\nu_{\rm R} =
U_{\rm R} (\hat{\omega}_{\rm L})^{\rm c} =
U_{\rm R} P_{\rm R} \chi,
\end{equation}
where $P_{\rm L,R} = (\mathbbm{1} \mp \gamma_5)/2$.

The unitarity of $\mathcal{U}$ is equivalent to the relations \cite{Grimus:2002ux}
\begin{equation}
\label{unitarity-relations}
U_{\rm L}^\dagger U_{\rm L} + U_{\rm R}^T U_{\rm R}^\ast = \mathbbm{1}_{n_{\rm L} + n_{\rm R}}
\quad \Leftrightarrow \quad
U_{\rm L}
U_{\rm L}^\dagger
=
\mathbbm{1}_{n_{\rm L}}, \, \,
U_{\rm R}
U_{\rm R}^\dagger
=
\mathbbm{1}_{n_{\rm R}}, \, \,
U_{\rm L}
U_{\rm R}^T
= 0_{n_{\rm L} \times n_{\rm R}}
\end{equation}
for the submatrices.
From (\ref{diag-prop}), written in the form $\mathcal{U}^T \mathcal{M}_\nu = \hat{\mathcal{M}}_\nu \, \mathcal{U}^\dagger$, one finds
\begin{eqnarray}
\label{rel1}
U_{\rm R}^\dagger M_{\rm D} =
\hat{\mathcal{M}}_\nu
U_{\rm L}^\dagger
\quad
&\Leftrightarrow&
\quad
M_{\rm D}^T
U_{\rm R}^\ast
=
U_{\rm L}^\ast
\hat{\mathcal{M}}_\nu,
 \label{diag-1} \\
U_{\rm L}^T
M_{\rm D}^T +
U_{\rm R}^\dagger
M_{\rm R}
=
\hat{\mathcal{M}}_\nu
U_{\rm R}^T
\quad
&\Leftrightarrow&
\quad
M_{\rm D}
U_{\rm L}
+
M_{\rm R}
U_{\rm R}^\ast
=
U_{\rm R}
\hat{\mathcal{M}}_\nu
.
\label{diag-2}
\end{eqnarray}
Combining the appropriate relations of (\ref{unitarity-relations}) and (\ref{diag-1}) implies
\begin{equation}
\label{relations}
U_{\rm L}^\ast
\hat{\mathcal{M}}_\nu
U_{\rm L}^\dagger = 0,
\quad
U_{\rm R}
\hat{\mathcal{M}}_\nu
U_{\rm L}^\dagger =
M_{\rm D}
\end{equation}
and
\begin{equation}
\label{M-nu-hat}
\hat{\mathcal{M}}_\nu
=
U_{\rm R}^\dagger M_{\rm D} U_{\rm L}
+
U_{\rm L}^T M_{\rm D}^T U_{\rm R}^\ast
+
U_{\rm R}^\dagger M_{\rm R} U_{\rm R}^\ast.
\end{equation}

Let us consider the diagonalization of the lowest-order neutrino mass matrix (\ref{M-nu}) in the 
interesting case 
$n_{\rm L} > n_{\rm R}$. 
Following the argument given in \cite{Grimus:1989pu}, 
the submatrix $M_{\rm D}$ can be regarded as a linear mapping $M_{\rm D} : 
\mathbbm{C}^{n_{\rm L}} 
\to \mathbbm{C}^{n_{\rm 
R}}$, where the relation
\begin{equation}
{\rm dim \, ker} \, M_{\rm D} = 
{\rm dim \, ker} \,  
\mathbbm{C}^{n_{\rm L}} 
-
{\rm dim \, im} \, M_{\rm D} 
\ge n_{\rm L} - n_{\rm R}
\end{equation}
ensures the existence of $n_{\rm L} - n_{\rm R}$ orthonormal vectors $u_i^\prime \in \mathbbm{C}^{n_{\rm L}}$ 
($1 \le i \le n_{\rm L} - n_{\rm R}$) with $M_{\rm D} u_i^\prime =0$. The unitary matrix $\mathcal{U}$ introduced in (\ref{intro-U}) can be written in 
the 
form
\begin{equation}
\mathcal{U} = \left( u_1, \ldots, u_{n_{\rm L} +n_{\rm R}} \right)
\end{equation}
with 
$n_{\rm L} + n_{\rm R}$ orthonormal vectors $u_i \in \mathbbm{C}^{n_{\rm L}+n_{\rm R}}$, where the first $n_{\rm L} - n_{\rm R}$ ones are given by
\begin{equation}
u_i =  
\begin{pmatrix} u_i^\prime \\
                  0          \\
                  \end{pmatrix}, \quad 1 \le i \le n_{\rm L} - n_{\rm R}.
\end{equation}
Thus, the decomposition
(\ref{decomp-1}) assumes the form \cite{Grimus:1989pu}
\begin{equation}
\label{decomp-2}
\mathcal{U} = \begin{pmatrix} U_{\rm L}^\prime & U_{\rm L}^{\prime \prime} \\
                                     0         & U_{\rm R}^{\prime \prime \, \ast} \\
              \end{pmatrix}
\end{equation}
with the $n_{\rm L} \times (n_{\rm L} - n_{\rm R})$ submatrix
$U_{\rm L}^\prime = (u_1^\prime, \ldots, u_{n_{\rm L} - n_{\rm R}}^\prime)$, the $n_{\rm L} \times 2 n_{\rm R}$ submatrix $U_{\rm L}^{\prime \prime}$
and the $n_{\rm R} \times 2 n_{\rm R}$ submatrix $U_{\rm R}^{\prime \prime \, \ast}$. 
Using (\ref{decomp-2}) in (\ref{diag-prop}),
we finally arrive at 
\begin{equation}
\label{diag-M}
\hat{\mathcal{M}}_\nu = \begin{pmatrix} 0 & 0 \\
                                            0 & \hat{M}^\prime \\
                  \end{pmatrix}
\end{equation}
with a diagonal and positive $2 n_{\rm R} \times 2 n_{\rm R}$ matrix 
$\hat{M}^\prime$. Thus, at tree level, there are in general $n_{\rm L} - n_{\rm R}$  massless Majorana neutrinos and $2 n_{\rm R}$ 
massive ones. Assuming a mass hierarchy $m_{\rm D} \ll m_{\rm R}$ in (\ref{M-nu}), the seesaw mechanism leads to $n_{\rm R}$ heavy neutral fermions 
and $n_{\rm R}$ light massive neutrinos.

\section{Interaction terms}
\label{sec: Interaction terms}
\renewcommand{\theequation}{\arabic{section}.\arabic{equation}}
\setcounter{equation}{0}

In this section we list the relevant interaction Lagrangians expressed in terms of mass eigenfields. 
For our present purposes, only the weak gauge couplings of the fermions and the Yukawa terms 
will be needed.

\subsection{Gauge couplings}
\label{subsec: Gauge couplings}

The fermionic gauge interactions are readily obtained from (\ref{Lfermion}) by rewriting the covariant derivative in terms of the mass 
eigenfields of the 
vector bosons. Omitting the gluonic term, (\ref{cov der}) takes the form
\begin{equation}
\label{cov der mef}
D_\mu = \partial_\mu
+ \frac{ig}{\sqrt{2}} \left(T_+ W_\mu^+ + T_- W_\mu^- \right) +\frac{ig}{\cos \theta_{\rm W}} \left(T_3- \sin^2 \! \theta_{\rm W} Q_{\rm em} \right) Z_\mu + i e Q_{\rm em} A_\mu,
\end{equation}
where
$
T_\pm = T_1 \pm i T_2$ and $e= g \sin \theta_{\rm W}$.
In this manner, we obtain the
charged-current Lagrangian 
\begin{equation}
\mathcal{L}_{\rm cc} = - \frac{g W_\mu^+}{\sqrt{2}} \left(
\overline{\chi} U_{\rm L}^\dagger \gamma^\mu P_{\rm L} \ell
+\overline{u} \gamma^\mu V_{\rm CKM} P_{\rm L} d \right) + {\rm h.c.}
\end{equation}
and the neutral-current Lagrangian 
\begin{eqnarray}
\mathcal{L}_{\rm nc} &=& - \frac{g Z_\mu}{\cos \theta_{\rm W}} \left[ \frac{1}{4} \overline{\chi} \gamma^\mu \left( U_{\rm L}^\dagger U_{\rm L} P_{\rm 
L} -U_{\rm L}^T U_{\rm L}^\ast P_{\rm R} \right) \chi
+ \overline{\ell} \gamma^\mu \left(
- \frac{1}{2} P_{\rm L} 
+ \sin^2 \! \theta_{\rm W} 
\right) \ell \right. \nonumber \\
&& + \left.
\overline{u} \gamma^\mu \left(\frac{1}{2} P_{\rm L} - \frac{2}{3} \sin^2 \! \theta_{\rm W} \right) u
+ \overline{d} \gamma^\mu \left(-\frac{1}{2} P_{\rm L} + \frac{1}{3} \sin^2 \! \theta_{\rm W} \right) d
\right].
\end{eqnarray}

\subsection{Yukawa couplings}
\label{subsec: Yukawa couplings}

The
Yukawa terms of the charged scalars ($1 \le a \le n_{\rm H}$) are given by
\begin{equation}
\label{Yuk-ch-sc}
\mathcal{L}_{\rm Y}
(H_a^\pm)
=
H_a^+ \left[ \overline{\chi} \left( \widetilde{Y}_{{\rm D} a} P_{\rm L} - \widetilde{Y}_a^{(\ell) \dagger} P_{\rm R} \right) \ell
+ \overline{u} \left(\widetilde{Y}^{(u)}_a P_{\rm L}- \widetilde{Y}_a^{(d) \dagger} P_{\rm R} \right) d \right]
+ {\rm h.c.},
\end{equation}
where we have introduced the coupling matrices
\begin{eqnarray}
\widetilde{Y}_{{\rm D} a} &=& U_{\rm R}^\dagger \sum\limits_{k=1}^{n_{\rm H}} U_{k a} Y_{{\rm D} k}, \qquad 
\widetilde{Y}^{(\ell)}_a =  \sum\limits_{k=1}^{n_{\rm H}} U_{k a}^\ast Y^{(\ell)}_k  U_{\rm L}, \\
\widetilde{Y}^{(u)}_a &=&   \sum\limits_{k=1}^{n_{\rm H}} U_{k a} Y^{(u)}_k  V_{\rm CKM}, \qquad
\widetilde{Y}^{(d)}_a =  \sum\limits_{k=1}^{n_{\rm H}} U_{k a}^\ast Y^{(d)}_k.
\end{eqnarray}

The Yukawa interactions of the neutral scalars ($0 \le b \le 2 n_{\rm H}$) take the form
\begin{eqnarray}
\label{Yuk-neutral-scalars}
\mathcal{L}_{\rm Y}(S^0_b) &=& - \frac{S_b^0}{\sqrt{2}}   \bigg[ \overline{\ell} \left(
\hat{Y}^{(\ell)}_b P_{\rm L}
+
\hat{Y}_b^{(\ell) \dagger} P_{\rm R} \right) \ell
+ \displaystyle\frac{1}{2} \overline{\chi}
\left(
F_b P_{\rm L}
+
F_b^\dagger P_{\rm R}
\right)
\chi
\nonumber \\
&& {} +
\overline{d}
\left(
\hat{Y}^{(d)}_b
V_{\rm CKM} P_{\rm L}
+
V_{\rm CKM}^\dagger
\hat{Y}^{(d) \dagger}_b
P_{\rm R}
\right)
d
+
\overline{u}
\left(
\hat{Y}^{(u)}_b  P_{\rm L}
+
\hat{Y}^{(u) \dagger}_b
P_{\rm R}
\right)
u
\bigg],
\end{eqnarray}
where
\begin{equation}
\label{Fb}
F_b =
U_{\rm R}^\dagger
\hat{Y}_{{\rm D} b}
U_{\rm L}
+
U_{\rm L}^T
\hat{Y}_{{\rm D} b}^T
U_{\rm R}^\ast
+
\sqrt{2} r_b
U_{\rm R}^\dagger
Y_{\rm R}
U_{\rm R}^\ast
\end{equation}
and
\begin{eqnarray}
\label{Y-hat}
\hat{Y}^{(\ell)}_b &=& \sum\limits_{k=1}^{n_{\rm H}} Y^{(\ell)}_k R_{k b}^\ast,
\qquad
\hat{Y}_{{\rm D} b} = \sum\limits_{k=1}^{n_{\rm H}} Y_{{\rm D} k} R_{k b},
\\
\hat{Y}^{(d)}_b &=& \sum\limits_{k=1}^{n_{\rm H}} Y^{(d)}_k R_{k b}^\ast,
\qquad
\hat{Y}^{(u)}_b = \sum\limits_{k=1}^{n_{\rm H}} Y^{(u)}_k R_{k b}.
\end{eqnarray}
Note the presence of the extra piece with $r_b$ and the additional scalon coupling (for $b=0$) in (\ref{Fb}) compared to the corresponding Yukawa 
couplings of the model described in \cite{Grimus:1989pu}.

\section{Scalon properties}
\label{sec: scalon}
\renewcommand{\theequation}{\arabic{section}.\arabic{equation}}
\setcounter{equation}{0}

The general (model-independent) features of the scalon interactions had already been noticed in \cite{Gildener:1976ih}. They follow from the 
fact that the fundamental order parameter $V$ and the scalon $S$ enter in the Lagrangian only via the combination $\phi = V +S$ parametrizing 
the flat direction of the tree potential.\footnote{See (\ref{ray}), (\ref{vacuum}), (\ref{def-rho-sigma}) and (\ref{rel-rho-sigma}) in the case of our 
model.}

As a consequence, the Yukawa couplings of the scalon are obtained by simply multiplying the fermion mass terms by the factor $S/V$,  
\begin{equation}
\label{Yuk-S}
\mathcal{L}_{\rm Y}(S) = - \frac{S}{V}
\left(
\overline{\ell} M_{\ell} \ell
+ \displaystyle\frac{1}{2}
\overline{\chi} \hat{\mathcal{M}}_{\nu} \chi
+
\overline{d} M_{d} d
+
\overline{u} M_{u} u
\right),
\end{equation}
which can also be checked by using (\ref{Yuk-neutral-scalars}-\ref{Y-hat}) for $b=0$ together with (\ref{M-l}-\ref{M-D}) and (\ref{M-nu-hat}). 

Similarly, the scalon couplings generated by (\ref{L-s}) are related to the mass terms of the vector bosons through the interaction Lagrangian 
\begin{equation}
\label{S-VV}
\left( M_W^2 W_\mu^+ W^{- \mu} + \frac{1}{2} M_Z^2 Z_\mu Z^\mu \right)
\left( \frac{2 S}{V} + \frac{S^2}{V^2} - \frac{2 n_0}{v^2} \! \sum\limits_{b=1}^{2 n_{\rm H} -1} \! \! \! r_b S^0_b   S\right),
\end{equation} 
where Goldstone boson terms have not been included.

The extraction of the scalon couplings from the scalar tree-potential is simplified by organizing
(\ref{treepotential}) in powers of $V+S$. Terms proportional to $(V+S)^4$ and $(V+S)^3$ are both absent because of the GW condition 
(\ref{GW-condition}) and the minimum condition (\ref{stat-cond}). The piece with $(V+S)^2$ contains the scalar mass terms together with 
the 
associated scalon couplings. Therefore, the latter take the form
\begin{equation}
\label{S-HH}
- \left( \sum\limits_{a=1}^{n_{\rm H}-1} \! \mu_a^2 H_a^+ H_a^- + \frac{1}{2} \sum\limits_{b=1}^{2 n_{\rm H} -1} \! \! M_b^2 (S_b^0)^2 \right) 
\left(\frac{2 
S}{V} + \frac{S^2}{V^2} \right).
\end{equation}
Finally, the term proportional to $V+S$ relates cubic scalar couplings (without the scalon) to quartic scalar couplings with exactly one scalon. As 
they 
will not be needed in the following, we refrain from displaying them here.

A glance at (\ref{Yuk-S}-\ref{S-HH}) reveals that the scalon decays predominantly into a pair of the heaviest possible states \cite{Gildener:1976ih}.
There remains the question whether the state $S$ can be identified with the $H^0$ particle discovered at a mass of $125 \, \rm GeV$ at the LHC 
\cite{Aad:2012tfa,Chatrchyan:2012xdj}. Studies of the $H^0$ decay properties performed so far are consistent with the SM predictions for the Higgs 
particle. The  
$H^0$ signal strength (for combined final states) of $1.10 \pm 0.11$ \cite{Patrignani:2016xqp} can serve as a quantitative measure of  
possible (small) deviations from the SM. Comparison of the SM Higgs couplings with (\ref{Yuk-S}) and (\ref{S-VV}) shows that $S \equiv H^0$ would 
actually correspond to $V \simeq v$. However, as we envisage to implement the seesaw mechanism in our model, the necessary mass hierarchy in 
(\ref{M-D}) requires $|v_k| \ll v_0$ or, equivalently, $v \ll V$. Thus, in our preferred scenario, the identification of the scalon with the 
scalar boson observed at $125 \, \rm GeV$ is excluded and one of the  
states $S^0_b$ ($1 \le b \le 2 n_{\rm H}-1$) has to play the role of the $H^0$ particle. 
 
In this way, we arrive at the prediction of a new neutral scalar $S$ with couplings (\ref{Yuk-S}-\ref{S-HH}) suppressed by a factor $v/V$ 
compared to the corresponding interactions of the SM Higgs $H^0$. Inserting the contributions of the relevant states in 
(\ref{scalon-mass-general}), the 
scalon mass formula of our model reads 
\begin{equation}
\label{M-S-model}
M_S^2  = \frac{1}{8 \pi^2 V^2} 
\left( M_{H^0}^4 + 3 M_Z^4 + 6 M_W^4  - 12 m_t^4 + \! \! \! \sum\limits_{S^0_b \ne H^0} \! \! \! M_b^4 
+ 2 \sum\limits_a  \mu_a^4
- 2 \, {\rm Tr} \, \hat{\mathcal{M}}_\nu^4 \right),
\end{equation}
where the tiny contributions from quarks lighter than the top have not been included. 
The neutrino term is practically exclusively determined by the masses of the heavy neutral fermions with 
${\rm Tr} \, \hat{\mathcal{M}}_\nu^4 \simeq {\rm Tr} \, (M_{\rm R}^\ast M_{\rm R})^2$.
Compared to the masses of the heavy Higgs and fermion states, the scalon mass is suppressed by the loop factor present in 
(\ref{M-S-model}). 
A further 
suppression occurs through the (partial) cancellation of bosonic and fermionic contributions in 
(\ref{M-S-model}), still allowing a wide range of possible scalon mass values.  

The positivity condition $M_S^2 > 0$ leads to the mass inequality
\begin{equation}
\label{mass-inequ}
\sum\limits_{S^0_b \ne H^0} \! \! \! M_b^4 
+ 2 \sum\limits_a  \mu_a^4
> 
\underbrace{-M_{H^0}^4 - 3 M_Z^4 - 6 M_W^4  + 12 m_t^4}_{\simeq (317 \, {\rm GeV})^4} 
+ 2 \, {\rm Tr} \, \hat{\mathcal{M}}_\nu^4 ,
\end{equation}
which can only be fulfilled by sufficiently large masses of the additional Higgs fields compensating the contribution of the heavy neutrinos.

\section{Neutrino masses at one-loop order}
\label{sec: Neutrino masses at one-loop}
\renewcommand{\theequation}{\arabic{section}.\arabic{equation}}
\setcounter{equation}{0}

The one-loop corrections generate an $n_{\rm L} \times n_{\rm L}$  submatrix $\delta M_{\rm L}$ \cite{Grimus:1989pu}
in the upper left corner of the tree-level mass matrix (\ref{M-nu}), where at tree level there is a zero submatrix.
As a consequence, the sum of all one-loop graphs contributing to $\delta M_{\rm L}$ must be finite and the final result can be expressed in terms of 
the 
(tree-level) parameters of the theory. The submatrices $M_{\rm D}$ and $M_{\rm R}$ also receive corrections, in this case from one-loop graphs as well 
as from 
counterterms \cite{Grimus:2002nk}. Altogether, $\mathcal{M}_{\nu}$ gets a shift
\begin{equation}
\label{M-nu-1}
\mathcal{M}_\nu  
\to
\mathcal{M}_\nu^{(1)}  
=
\mathcal{M}_\nu +
\delta \mathcal{M}_\nu, \quad
\delta \mathcal{M}_\nu = \begin{pmatrix} \delta M_{\rm L} & \delta M_{\rm D}^T \\ 
                                         \delta M_{\rm D} & \delta M_{\rm R} \\
                         \end{pmatrix}.
\end{equation}

As we are only interested in the masses of those neutrino states that are massless at tree level, it is sufficient to compute the 
submatrix $\delta M_{\rm L}$. To this end, we consider the 
neutrino self-energy matrix in the basis of the tree-level mass eigenfields. We use the decomposition 
\cite{Grimus:2002nk}
\begin{equation}
\Sigma(p) =
A_{\rm L} (p^2) \slashed{p} P_{\rm L}
+
A_{\rm R} (p^2) \slashed{p} P_{\rm R}
+
B_{\rm L} (p^2) P_{\rm L}
+
B_{\rm R} (p^2) P_{\rm R},
\end{equation}
where $p$ is the neutrino four-momentum. The
dispersive parts of the coefficients satisfy
\begin{equation}
A_{\rm L}^\dagger = A_{\rm L}, \quad
A_{\rm R}^\dagger = A_{\rm R}, \quad
B_{\rm L}^\dagger = B_{\rm R}.
\end{equation}
The Majorana nature 
$\chi^{\rm c} = \chi$ of the neutrino field implies the consistency condition
\begin{equation}
\Sigma(p) = C \Sigma(-p)^T C^{-1} \quad \Rightarrow \quad
A_{\rm L} = A_{\rm R}^T, \quad
B_{\rm L} = B_{\rm L}^T, \quad
B_{\rm R} = B_{\rm R}^T.
\end{equation}

For our purposes, it suffices to consider only $B_{\rm L}(p^2)$ at $p^2=0$. Transforming back to the original basis one arrives at
\cite{Grimus:2002nk}
\begin{equation}
\label{delta-M-L}
\delta M_{\rm L} = 
U_{\rm L}^\ast 
B_{\rm L} (0) 
U_{\rm L}^\dagger 
=
U_{\rm L}^\ast 
\left[
B_{\rm L}^{(Z^0)} (0)  
+ \sum\limits_{b=0}^{2 n_{\rm H}} B_{\rm L}^{(S_b^0)} (0)  
\right]
U_{\rm L}^\dagger, 
\end{equation}
receiving contributions from one-loop graphs with the $Z^0$ boson and neutral scalars. 
Following  
\cite{Grimus:2002nk} we compute (\ref{delta-M-L})
in a general $R_\xi$ gauge. $B_{\rm L}^{(Z^0)}(0)$ turns out to be finite, once the relation $U_{\rm L}^\ast 
\hat{\mathcal{M}}_\nu U_{\rm L}^\dagger = 0$ from (\ref{relations}) is used. Employing also $U_{\rm L} U_{\rm L}^\dagger = \mathbbm{1}_{n_{\rm 
H}}$ 
displayed in (\ref{unitarity-relations}), one obtains the (gauge dependent) expression  
\begin{eqnarray}
\label{delta-M-Z} 
U_{\rm L}^\ast 
B_{\rm L}^{(Z^0)} (0) 
U_{\rm L}^\dagger 
\!
&=& 
\!
\frac{M_Z^2}{(4 \pi)^2  v^2} 
U_{\rm L}^\ast  
\left\{ 4 \hat{\mathcal{M}}_\nu  \int\limits_0^1 \! d\alpha \, 
\ln \left( \alpha M_Z^2 +(1-\alpha) \hat{\mathcal{M}}_\nu^2 \right) 
\right.
 \\
\!
&+& 
\!
\left.  \frac{\hat{\mathcal{M}}_\nu^3}{M_Z^2}
\int\limits_0^1 \! d \alpha \left[ \ln \left( \alpha \xi_Z M_Z^2 + (1-\alpha) \hat{\mathcal{M}}_\nu^2 \right)
-
\ln \left( \alpha  M_Z^2 + (1-\alpha) \hat{\mathcal{M}}_\nu^2 \right) \right] \right\}
U_{\rm L}^\dagger . \nonumber
\end{eqnarray}

In the scalar contribution to (\ref{delta-M-L}) we encounter the products $U_{\rm L}^\ast F_b$ and $F_b U_{\rm L}^\dagger$, respectively. Using the 
unitarity relations (\ref{unitarity-relations}), these terms can be rewritten as
\begin{equation} 
U_{\rm L}^\ast F_b = \hat{Y}_{{\rm D} b}^T U_{\rm R}^\ast, \quad 
F_b U_{\rm L}^\dagger =
U_{\rm R}^\dagger \hat{Y}_{{\rm D} b},
\end{equation}
which shows that the additional pieces proportional to $r_b$, present in (\ref{Fb}), do not appear in (\ref{delta-M-L}). The sum of all 
scalar loops is finite because 
of the last relation in (\ref{orth-rel}) and we obtain 
\begin{equation}
\label{scalar-sum}
U_{\rm L}^\ast 
\sum\limits_{b=0}^{2 n_{\rm H}} B_{\rm L}^{(S_b^0)} (0)  
U_{\rm L}^\dagger
= \frac{1}{2 (4 \pi)^2} 
\sum\limits_{b=0}^{2 n_{\rm H}} 
\hat{Y}_{{\rm D} b}^T U_{\rm R}^\ast
\hat{\mathcal{M}}_\nu
\int\limits_0^1 \! d \alpha \ln \left( \alpha  M_b^2 + (1-\alpha) \hat{\mathcal{M}}_\nu^2 \right)
U_{\rm R}^\dagger \hat{Y}_{{\rm D} b}.
\end{equation}

Using $\hat{Y}_{{\rm D} \, 2n_{\rm H}} = i \sqrt{2} M_{\rm D}/v$ and (\ref{rel1}), the contribution of the neutral Goldstone boson 
$G^0 \equiv S^0_{2 n_{\rm H}}$ 
can be recast into the form
\begin{equation}
\label{delta-M-G0}
-\frac{1}{(4 \pi)^2 v^2} U_{\rm L}^\ast \hat{\mathcal{M}}_\nu^3
\int\limits_0^1 \! d \alpha \ln \left( \alpha \xi_Z M_Z^2 + (1-\alpha) \hat{\mathcal{M}}_\nu^2 \right)
U_{\rm L}^\dagger,
\end{equation}
and we see that this term cancels exactly the
gauge-dependent part
of (\ref{delta-M-Z}). 

The total result for $\delta M_{\rm L}$ is now given by
\begin{eqnarray}
\label{delta-M-L-tot}
\delta M_{\rm L} 
&=& 
\frac{3}{(4 \pi)^2 v^2} 
U_{\rm L}^\ast 
\hat{\mathcal{M}}_\nu^3 
\frac{\ln (
\hat{\mathcal{M}}_\nu^2/M_Z^2
)}{\hat{\mathcal{M}}_\nu^2/M_Z^2-1}
U_{\rm L}^\dagger
\nonumber
\\
&+& 
\frac{1}{(4 \pi )^2 V^2} 
U_{\rm L}^\ast 
\hat{\mathcal{M}}_\nu^3 
\frac{\ln (
\hat{\mathcal{M}}_\nu^2/M_S^2
)}{\hat{\mathcal{M}}_\nu^2/M_S^2-1}
U_{\rm L}^\dagger
\nonumber
\\
&+&
\frac{1}{2 (4 \pi)^2} 
\sum\limits_{b=1}^{2 n_{\rm H}-1} 
\hat{Y}_{{\rm D} b}^T U_{\rm R}^\ast
\hat{\mathcal{M}}_\nu 
\frac{\ln (
\hat{\mathcal{M}}_\nu^2/M_b^2
)}{\hat{\mathcal{M}}_\nu^2/M_b^2-1}
U_{\rm R}^\dagger
\hat{Y}_{{\rm D} b} .
\end{eqnarray}
The first term is the residual $Z^0$ contribution once all cancellations have been taken into account.  The second one represents the scalon 
contribution 
$\delta M_{\rm L}^{(S)}$ 
obtained 
from the summand with $b=0$ in (\ref{scalar-sum})  
by using 
$\hat{Y}_{{\rm D} 0} =  \sqrt{2} M_{\rm D}/V$ and (\ref{rel1}).
The third term contains the contributions of all physical scalars except the scalon.

Recalling (\ref{Y-hat}), the Yukawa matrices in 
(\ref{delta-M-L-tot})
are 
related to the ones introduced in (\ref{R-L-Yuk})  by $\hat{Y}_{{\rm D} b} = \sum_k Y_{{\rm D} k} R_{k b}$. The orthogonality relation $R R^T 
= 0$ of the 
transformation matrix $R$
given in (\ref{orth-rel}) can be rewritten as
\begin{equation}
\label{mod-rel}
\sum\limits_{b=1}^{2 n_{\rm H}-1} \! R_{kb} R_{\ell b} = \frac{v_k v_\ell}{v^2} \left( 1 - \frac{v^2}{V^2} \right), 
\end{equation}
modifying the corresponding formula in the original version of the model \cite{Grimus:1989pu,Grimus:1999wm} only by the extra factor 
$1-v^2 /V^2 \simeq 1$.

Strictly speaking, the scalon contribution to $\delta M_{\rm L}$ vanishes at the order we are working.  
To be fully consistent at one-loop, we have to insert tree-level masses in (\ref{delta-M-L-tot}) corresponding to $M_S = 0$ in the case of the scalon, 
which leads  
to $\delta M_{\rm L}^{(S)} = 0$ in this limit. In this sense and apart from the small modification in (\ref{mod-rel}), we recover the 
same one-loop result for $\delta M_{\rm L}$ that had already been obtained in the non-conformal model without additional scalar 
singlet \cite{Grimus:2002nk}.

The diagonalization of the one-loop mass matrix $\mathcal{M}_\nu^{(1)}$ was described in \cite{Grimus:1989pu}. As a net result, the diagonal 
tree-level mass matrix (\ref{diag-M}) with zero entries in the left upper block is replaced by the one-loop matrix 
\begin{equation}
\hat{\mathcal{M}}_\nu^{(1)}
= \begin{pmatrix} \hat{M}_0 & 0 \\ 
                           0 & \hat{M}^\prime \\
                         \end{pmatrix}, \qquad 
\hat{M}_0 = 
U_{\rm L}^{\prime \, T} \delta M_{\rm L} U_{\rm L}^\prime,
\end{equation}
where the $n_{\rm L} - n_{\rm R}$ orthonormal vectors in 
$U_{\rm L}^\prime = (u_1^\prime, \ldots, u_{n_{\rm L}-n_{\rm R}}^\prime)$, 
so far only subject to the constraint $M_{\rm D} u_i^\prime = 0$, 
can now be chosen in such a way that 
$\hat{M}_0$ 
becomes diagonal and non-negative.  
 
The $Z^0$ as well as the scalon contribution in (\ref{delta-M-L-tot}) have the general structure
\begin{equation}
U_{\rm L}^\ast 
\hat{\mathcal{M}}_\nu
f(\hat{\mathcal{M}}_\nu)
\hat{\mathcal{M}}_\nu
U_{\rm L}^\dagger
= 
M_{\rm D}^T U_{\rm R}^\ast
f(\hat{\mathcal{M}}_\nu)
U_{\rm R}^\dagger
M_{\rm D},
\end{equation}
where the form on the right-hand side follows from (\ref{rel1}). Because of $M_{\rm D} U_{\rm L}^\prime = 0$, both terms do not contribute to the 
one-loop induced 
neutrino 
masses in 
$\hat{M}_0$. 
The explicit expression is given by
\begin{equation}
\label{M-0}
\hat{M}_0 
=
\frac{1}{2 (4 \pi)^2} 
U_{\rm L}^{\prime \, T} 
\left(
\sum\limits_{b=1}^{2 n_{\rm H}-1} 
\hat{Y}_{{\rm D} b}^T U_{\rm R}^\ast
\hat{\mathcal{M}}_\nu 
\frac{\ln (
\hat{\mathcal{M}}_\nu^2/M_b^2
)}{\hat{\mathcal{M}}_\nu^2/M_b^2-1}
U_{\rm R}^\dagger
\hat{Y}_{{\rm D} b} 
\right)
U_{\rm L}^\prime .
\end{equation}
Neglecting terms suppressed by a factor $m_{\rm D} / m_{\rm R}$, (\ref{M-0}) can be rewritten as
\begin{equation}
\hat{M}_0 
\simeq
\frac{1}{2 (4 \pi)^2} 
U_{\rm L}^{\prime \, T} 
\left(
\sum\limits_{b=1}^{2 n_{\rm H}-1} 
\hat{Y}_{{\rm D} b}^T 
M_{\rm R}^\ast
\frac{\ln (
M_{\rm R} M_{\rm R}^\ast /M_b^2
)}{M_{\rm R} M_{\rm R}^\ast/M_b^2-1}
\hat{Y}_{{\rm D} b} 
\right)
U_{\rm L}^\prime ,
\end{equation}
which is just the form of the result given in formula (3.10) of \cite{Grimus:1989pu}.

Hence, concerning the neutrino mass spectrum, we encounter essentially the same situation as in the case of the conventional multi-Higgs-doublet model 
with right-handed neutrinos.
In general (i.e. without any relations among the Yukawa matrices $\hat{Y}_{{\rm D} b}$),
\begin{equation}
\nu_0 = \max \left( 0, n_{\rm L} - n_{\rm H} n_{\rm R} \right)
\end{equation}
of the $n_{\rm L} -n_{\rm R}$ neutrinos with vanishing tree-level masses remain massless at the one-loop level \cite{Grimus:1989pu}, whereas the 
other $n_{\rm L} - n_{\rm R} - \nu_0$ states receive masses from (\ref{M-0}). 

\section{Conclusions}
\label{sec: Conclusions}
\renewcommand{\theequation}{\arabic{section}.\arabic{equation}}
\setcounter{equation}{0}

The purpose of this work was the construction of a conformal ${\rm SU}(2)_{\rm L} \times {\rm U}(1)_Y$ 
multi-Higgs-doublet model being able to reproduce the attractive features of the SM extension proposed in 
\cite{Grimus:1989pu}, where the neutrino mass spectrum results from a combination of the seesaw mechanism at the tree-level with higher-order mass 
production. Starting from the particle content described in \cite{Grimus:1989pu} 
(with $n_{\rm L}$ generations of SM fermions,
$n_{\rm R}$ right-handed neutrino gauge-singlets and
$n_{\rm H}$ Higgs doublets), already the simplest possible enlargement of the scalar sector by only one real singlet field turned out to be sufficient 
to 
attain 
the goal.  

The conformal symmetry of the model allows only terms with (operator) dimension four in the Lagrangian. Therefore, the associated coupling parameters 
must be dimensionless and, in contrast to the model suggested in \cite{Grimus:1989pu}, explicit mass terms for the right-handed fermion singlets or 
the scalars as well as trilinear scalar-couplings are forbidden.
However, already at the one-loop level, the scale invariance of the classical theory is broken by quantum corrections: 
the CW mechanism generates a mass 
scale by dimensional transmutation \cite{Coleman:1973jx}. The order parameter $V$, which determines the characteristic scale of the model, is the 
common origin of all dimensionful quantities in the EW sector, including all vacuum expectation values 
(in particular the EW scale $v \simeq 
246 \, {\rm GeV}$) and a fortiori of  
the masses of quarks and leptons.
In this framework, the EW 
interaction is put on an equal footing with QCD, where the characteristic scale $\Lambda_{\rm QCD}$ of the strong interaction is also a consequence of 
dimensional transmutation.            

The seesaw mechanism 
\cite{Minkowski:1977sc,Yanagida:1979as,Glashow:1979nm, GellMann:1980vs,Mohapatra:1979ia}
is an essential ingredient of our model. In the absence of explicit mass terms in the Lagrangian, 
the Yukawa interaction of the scalar singlet with the right-handed neutrino singlets produces a Majorana mass term at a scale $m_{\rm R} \sim 
V$, once the scalar field receives its vacuum expectation value through the CW mechanism. 
At the same time, the Yukawa terms of the Higgs doublets produce Dirac mass terms at a scale $m_{\rm D}$, where 
the necessary mass hierarchy $m_{\rm D} \ll m_{\rm R}$ (being in general equivalent to $v \ll V$) 
can be accommodated (though not ``explained") in the  framework of this model. One encounters an analogous situation in the description of the mass 
spectra of quarks and charged leptons: the observed mass values serve as input parameters in some Yukawa coupling matrices, but cannot be deduced from 
the 
theory.

The scalar spectrum of the model comprises $n_{\rm H} -1$ charged physical scalars and $2 n_{\rm H}$ neutral ones. Two of the neutral scalars stand out 
by their specific properties: the scalon $S$ \cite{Gildener:1976ih} and the Higgs particle $H^0$ observed at $125 \, \rm GeV$ 
\cite{Aad:2012tfa,Chatrchyan:2012xdj}. The scalon becomes massive at one-loop order, where its mass  
can be expressed in terms of the masses of all other fields. The scalon couplings to the other particles can be derived in a  
model-independent way \cite{Gildener:1976ih}. They are flavour-diagonal but 
differ by a factor $v/V$ from the corresponding interactions of the SM Higgs field. This observation together with the experimentally measured decay 
rates of the $H^0$ 
\cite{Patrignani:2016xqp}, being in good agreement with the predictions of the SM,
inhibit the identification of the scalon with the $H^0$ particle if $v/V \ll 1$ is assumed.
Thus, the existence  of a particle with the properties of the scalon is a genuine prediction of our model. 

The positivity condition for the squared scalon mass (ensuring also the stability of the theory) leads to a mass inequality, which can only be 
satisfied by 
sufficiently heavy scalar fields with masses of comparable size to those of the heavy neutrinos, suggesting a natural reason for the suppression of 
flavour-changing neutral interactions by neutral scalar exchange.
   
Compared to the heavy masses (typically of the order $V$), the scalon mass is suppressed by a one-loop factor and, in addition,
by opposite signs of  bosonic and fermionic contributions in its mass formula. Still, a wide range of possible scalon mass values (with quite 
different phenomenological scenarios) remains open.

An important part of this work was dedicated to the exploration of the neutrino mass spectrum in the framework of the conformal 
theory. Extending and adapting the calculation performed in \cite{Grimus:2002nk} for our purposes, we have computed the dominant one-loop corrections 
to the 
tree-level mass matrix of the light neutrinos. We have explicitly checked the finiteness and the gauge independence (in a general $R_\xi$ gauge) of 
these contributions. Applying our results to the interesting case $n_{\rm L} > n_{\rm R}$, we recover the same features of the neutrino 
mass spectrum as in 
\cite{Grimus:1989pu}, demonstrating that also the conformal version of the model is capable of combining the seesaw mechanism with higher-order mass 
production.   
Assuming a sufficiently high scale $V \gg v$, there are $n_{\rm R}$ heavy neutral leptons, $n_{\rm R}$ neutrinos being light because of the seesaw 
mechanism and $n_{\rm L} - n_{\rm R}$ massless neutrinos at the tree level. At the one-loop level, 
$\nu_0 = \max (0, n_{\rm L} -n_{\rm H} n_{\rm R})$ states stay massless, whereas the remaining
$n_{\rm L} - n_{\rm R} - \nu_0$ ones become massive with naturally small and calculable masses generated by the exchange of neutral physical scalars, 
where the scalon contribution was shown to vanish to the order considered. As already emphasized in \cite{Grimus:1999wm},
the dominant one-loop corrections to the seesaw mechanism are quadratic in the Yukawa couplings (just like the
masses of the light neutrinos at tree-level) and the one-loop corrections are smaller than the tree-level results only because of
the appearance of the loop factor $1/ (4 \pi)^2$.

Finally, the simplest non-trivial version of the model with just two Higgs doublets and a single right-handed 
neutrino ($n_{\rm H} = 2$, $n_{\rm R} =1$) may serve as an illustration of our findings.\footnote{A detailed phenomenological analysis of this minimal 
version of the model is in 
preparation.}
In this case, the spectrum of the theory contains 
four Majorana masses with a natural
hierarchy. One heavy and one light mass are obtained by the seesaw mechanism, the next one appears at the one-loop level and the last one at two
loops. The scalar sector consists of a single charged Higgs field $H^\pm$ and four neutral ones ($H^0$, $S$, $H_1^0$, $H_2^0$). The scalon mass formula 
takes the form
\begin{equation}
M_S^2  = \frac{1}{8 \pi^2 V^2} 
\left( M_{H^0}^4 + 3 M_Z^4 + 6 M_W^4  - 12 m_t^4 + 
M_{H_1^0}^4 
+M_{H_2^0}^4 
+ 2 M_{H^\pm}^4 
- 2  M_{\rm R}^4 \right),
\end{equation}
where
the positivity condition $M_S^2 > 0$ is equivalent to the mass inequality
\begin{equation}
\label{mass-inequality}
M_{H_1^0}^4 
+M_{H_2^0}^4 
+ 2 M_{H^\pm}^4 
> 
\underbrace{-M_{H^0}^4 - 3 M_Z^4 - 6 M_W^4  + 12 m_t^4}_{\simeq (317 \, {\rm GeV})^4} 
+ \, 2 M_{\rm R}^4. 
\end{equation}
As a consequence, the large 
Majorana mass 
contribution 
on the right-hand side of 
(\ref{mass-inequality}) 
requires 
sufficiently heavy states 
in the scalar sector.

\section*{Acknowledgements}
Discussion with Michael Lansch\"{u}tzer and Maximilian L\"{o}schner are gratefully acknowledged. We thank Walter Grimus and Gerhard Ecker for reading 
the 
manuscript and valuable comments.



\begin{thebibliography}{99}

\bibitem{Grimus:1989pu}
  W.~Grimus and H.~Neufeld,
  \emph{Radiative neutrino masses in an $ SU(2) \times U(1)$ model},
  \emph{Nucl. Phys.} {\bf B  325} (1989) 18.

\bibitem{Grimus:2002ux}
  W.~Grimus and L.~Lavoura,
  \emph{Soft lepton-flavor violation in a multi-Higgs-doublet seesaw model},
  \emph{Phys. Rev.} {\bf D 66} (2002) 014016,
  arXiv:hep-ph/0204070.

\bibitem{Minkowski:1977sc}
  P.~Minkowski,
  \emph{$\mu \to e\gamma$ at a rate of one out of $10^{9}$ muon decays?},
  \emph{Phys. Lett.} {\bf B 67} (1977) 421.

\bibitem{Yanagida:1979as}
  T.~Yanagida,
  \emph{Horizontal symmetry and masses of neutrinos},
  in: O.~Sawata, A.~Sugamoto (eds.), Proceedings of the Workshop on Unified Theory and Baryon
  Number in the Universe, Tsukuba, Japan, 1979, KEK report 79-18.

\bibitem{Glashow:1979nm}
  S.~L.~Glashow,
  \emph{The future of elementary particle physics},
  in: M.~L\'{e}vy et al. (eds.), Quarks and Leptons, Proceedings of the Advanced Study  Institute, Carg\`{e}se
  Corsica, 1979, Plenum, New York, 1980.

\bibitem{GellMann:1980vs}
  M.~Gell-Mann, P.~Ramond and  R.~Slansky,
  \emph{Complex spinors and unified theories},
  in: D.Z.~Freedmann, F.~van~Nieuwenhuizen, (eds.), Supergravity, North Holland, Amsterdam, 1979,
  arXiv:1306.4669 [hep-th].

\bibitem{Mohapatra:1979ia}
  R.~N.~Mohapatra and G.~Senjanovi\'{c},
  \emph{Neutrino mass and spontaneous parity violation},
  \emph{Phys. Rev. Lett.} {\bf  44} (1980) 912.

\bibitem{Grimus:2002nk}
  W.~Grimus and L.~Lavoura,
  \emph{One-loop corrections to the seesaw mechanism in the multi-Higgs-doublet standard model},
  \emph{Phys. Lett.} {\bf B 546} (2002) 86,
  arXiv:hep-ph/0207229.

\bibitem{Grimus:1999wm}
  W.~Grimus and H.~Neufeld,
  \emph{3-neutrino mass spectrum from combining seesaw and radiative neutrino mass mechanisms},
  \emph{Phys. Lett.} {\bf B 486} (2000) 385,
  arXiv:hep-ph/9911465.

\bibitem{Ibarra:2011gn}
  A.~Ibarra and C.~Simonetto,
  \emph{Understanding neutrino properties from decoupling right-handed neutrinos and extra Higgs doublets},
  \emph{JHEP} {\bf 1111} (2011) 022,
  arXiv:1107.2386 [hep-ph].

\bibitem{Bardeen:1995kv}
  W.~A.~Bardeen,
  \emph{On naturalness in the standard model},
  FERMILAB-CONF-95-391-T (1995).

\bibitem{Coleman:1973jx}
  S.~Coleman and E.~Weinberg,
  \emph{Radiative corrections as the origin of spontaneous symmetry breaking},
  \emph{Phys. Rev.} {\bf D 7} (1973) 1888.

\bibitem{Goldstone:1962es}
  J.~Goldstone, A.~Salam and S.~Weinberg,
  \emph{Broken symmetries},
  \emph{Phys. Rev.} {\bf 127} (1962) 965.

\bibitem{JonaLasinio:1964cw}
  G.~Jona-Lasinio,
  \emph{Relativistic field theories with symmetry breaking solutions},
  \emph{Nuovo Cim.} {\bf 34} (1964) 1790.

\bibitem{Gildener:1976ih}
  E.~Gildener and S.~Weinberg,
  \emph{Symmetry breaking and scalar bosons},
  \emph{Phys. Rev.} {\bf D 13} (1976) 3333.

\bibitem{Helmboldt:2016mpi}
  A.~J.~Helmboldt, P.~Humbert, M.~Lindner and J.~Smirnov,
  \emph{Minimal conformal extensions of the Higgs sector},
  \emph{JHEP} {\bf 1707} (2017) 113,
  arXiv:1603.03603 [hep-ph].

\bibitem{Lindner:2014oea}
  M.~Lindner, S.~Schmidt and J.~Smirnov,
  \emph{Neutrino masses and conformal electro-weak symmetry breaking},
  \emph{JHEP} {\bf 1410} (2014) 177,
  arXiv:1405.6204 [hep-ph].

\bibitem{Weinberg:1973ua}
  S.~Weinberg,
  \emph{Perturbative calculations of symmetry breaking},
  \emph{Phys. Rev.} {\bf D 7} (1973) 2887.

\bibitem{Espinosa:2007qk}
  J.~R.~Espinosa and M.~Quiros,
  \emph{Novel effects in electroweak breaking from a hidden sector},
  \emph{Phys. Rev.} {\bf D 76} (2007) 076004,
  arXiv:hep-ph/0701145.

\bibitem{Cabibbo:1963yz}
  N.~Cabibbo,
  \emph{Unitary symmetry and leptonic decays},
  \emph{Phys. Rev. Lett.} {\bf 10} (1963) 531.

\bibitem{Kobayashi:1973fv}
  M.~Kobayashi and T.~Maskawa,
  \emph{CP violation in the renormalizable theory of weak interaction},
  \emph{Prog. Theor. Phys.} {\bf 49} (1973) 652.

\bibitem{Aad:2012tfa}
  G.~Aad et al. (ATLAS Collaboration),
  \emph{Observation of a new particle in the search for the Standard Model Higgs boson with the ATLAS detector at the LHC},
  \emph{Phys. Lett.} {\bf B 716} (2012) 1,
  arXiv:1207.7214 [hep-ex].

\bibitem{Chatrchyan:2012xdj}
  S.~Chatrchyan et al. (CMS Collaboration),
  \emph{Observation of a new boson at a mass of 125 GeV with the CMS experiment at the LHC},
  \emph{Phys. Lett.} {\bf B 716} (2012) 30,
  arXiv:1207.7235 [hep-ex].

\bibitem{Patrignani:2016xqp}
  C.~Patrignani et al. (Particle Data Group),
  \emph{Review of Particle Physics},
  \emph{Chin. Phys.} {\bf C 40} (2016) 100001.





\end{thebibliography}
\end{document}